# The planar Lanchester model of insurgent warfare: Intricate Collateral Damage Functions and Global Bifurcation

## Rouzbeh Aghaieebeiklavasani[1]

## Gholam Reza Rokni Lamouki[2]

**Abstract:**

One of the most notable aspects of mathematical modeling is that it sheds light on the complexities arising from changes in parameters and their real-world implications, thus gaining better insight into the dynamics of economic, political, and security phenomena. Moreover, modifications to mathematical modeling will set the stage for embedding new features into the system and enriching the assessment of our analysis. In the case of the Lanchester model of warfare, the introduction of the collateral damage function was an attempt to turn the model into a more realistic framework for understanding counter-insurgencies and irregular battles. In this article, we focus on addressing the impact of sophisticated collateral damage functions in a modified Lanchester model of combat, which reflects the multifaceted nature of warfare. This analysis shows the possibility of global bifurcations and their repercussions for the combat situation between two players. Finally, we assess the codimension two bifurcation analysis of our system based on the interaction between collateral damage and the effectiveness of targeting insurgents.

## Introduction

The history of combat is fraught with complexities arising from factors such as ethnicity, religion, physical conditions, third-party interventions, and domestic barriers or facilitators. Changes in both qualitative and quantitative elements can shape conflicts and influence their resolution. A retrospective analysis of warfare throughout military and political history enables us to categorize the various variables that govern war. Factors such as the number of combatants and military equipment, economic power, logistical challenges, and technological advancements define the boundaries of battles. More importantly, there is a deep interconnection between the qualitative and quantitative variables of war, making their separate analysis insufficient. This explains certain inconsistencies in our broader understanding of combat. For instance, in the context of the war on terror, examining the strategic culture and narratives of terror groups can fuel uncertainty and complexity, further diverging from the optimal solutions proposed by modeling. Therefore, in addition to the merely quantitative measures we include in our models, observing the qualitative measures through the lens of quantitative assessments and mathematical objects will help us study a unified system of conflicts and its connection to national security.

[1] Rouzbeh Aghaieebeiklavasani graduated from the faculty of Law and Political Science and from the School of Mathematics, Statistics, and Computer Science at the University of Tehran. He is currently an Independent Researcher who applies mathematical models in political science and economics.( Corresponding author) Email: robert12t.robert@gmail.com

[2] Gholam Reza Rokni Lamouki is an Associate Professor at the School of Mathematics, Statistics, and Computer Science, College of Science, University of Tehran, who works on control theory and dynamical systems and their real-world applications. Email: rokni@ut.ac.ir

When it comes to the appraisal of wars in the contemporary world, the effect of collateral damage is a sine qua non. In addition to the lethality of novel military equipment, the emergence of new complex battle environments, with the importance of public opinion in some societies on the rise, the impact of religion and ethnicity in driving angry population, and interdependent regional and international relations and diplomacy all give credence to the plausibility of this functional and its effect on the dynamics of the model like the emergence of the ups and downs in the dominance of either players.

Notwithstanding the importance of studying collateral damage in the dynamics of wars, specifically in the Lanchester model of warfare that highlights the number of combatants, delving deeply into this concept as a general mathematical object with unique characteristics will provide us with a genuine opportunity to study the complexity of environments and players' traits in a given conflict. Physical conditions like the third party's sheltering insurgents and the impact of excessive violence on collateral damage necessitate us to observe the interaction between targeting insurgents and collateral damage functions through a different prism. Furthermore, there is a difference, yet pertinence, between logistical and ideological aspects of collateral damage, explaining which parts of recruitment are subject to change by external forces. In this system, in tandem with a well-known trend in the Lanchester model that brings up the thresholds for victory of one side, it gives rise to the real-world application of global bifurcation analysis as a new boundary for determining the winners and losers of an asymmetric conflict, and lays bare some distinctive features of the model proposed by us.

In this article, we assess more realistic versions of collateral damage in the modified version of the Lanchester model, which is central to gaining better insight into combats and their unintended consequences, like how an increase in firepower, under some circumstances, can lead the exerting player to lose the combat. By introducing a category of complex collateral damages, we assess their mathematical features and the possibility of multistability in our system, which provides readers with a new definition of victory/loss threshold. A geometric analysis of equilibrium points' stability will decode the conditions for the emergence of the cycle of tension between players and non-trivial boundaries. In the meantime, the change in the firepower of the government concurrently with the insurgents' attempts to mitigate damages can generate Cusp bifurcation in the system. Finally, we will mention the possibility of heterogeneity in combat modeling, given the presence of multiple players, and show how treating collateral damage as a good distributed among multiple insurgents can change the dynamics of our model and its geometric interpretation.

## The evolution of the Lanchester model and counter-insurgencies

Fortunately, there is no shortage of scholarship on the systematic approach to wars and conflicts. Nevertheless, it is worth noting that classifying vast sources is necessary to acquire a profound understanding of the interaction between dynamical systems and conflicts. More importantly, this retrospective analysis will help us detect theoretical and technical flaws in conventional approaches to the Lanchester model. In 1916, Lanchester studied mathematical objects to describe air warfare.[3] This has widespread implications for the study of dynamical systems and technical issues in the analysis of war. Consequently, there have been various extensions, modifications, and different interpretations.[4], which include the interaction between Lanchester and asymmetric warfare[5], ground combat[6] the interwoven nature of cyber and physical domains in the Lanchester combat model7, Lanchester and optimal control theory[8], joint operations like AirLand combats[9] , multilateral conflict in Lanchester[10],  PDE  and Lanchester[11], Lanchester model of firepower and intelligence[12]. With the idea of collateral damage as a double-edge sword effect in counter-insurgencies[13]. Moreover, we can turn to optimization to study combats in the presence of this function[14].

[3] Frederick William, Lanchester,. Aircraft in warfare: The dawn of the fourth arm. Constable limited, 1916.
[4] G, Kaimakamis. and N. B. Zographopoulos. "On a Lanchester Combat Model." in Nicholas J. Daras (Ed.). *Applications of mathematics and informatics in military science* (Vol. 71). Springer, pp 113-116
Washburn, A. R., & Kress, M. (2009). *Combat modeling* (Vol. 139). New York: Springer. PP 79-109

---

[5] Marvin B, Schaffer. "Lanchester models of guerrilla engagements." Operations Research 16.3 (Summer 1968): 457-488 , https://doi.org/10.1287/opre.16.3.457 ;

Moshe Kress and Niall J. MacKay. "Bits or shots in combat? The generalized Deitchman model of guerrilla warfare." Operations Research Letters 42.1 (January2014): 102-108 , https://doi.org/10.1016/j.orl.2013.08.004

Michael P, Atkinson, Alexander Gutfraind, ahnd Moshe Kress. "When do armed revolts succeed: lessons from Lanchester theory." Journal of the Operational Research Society 63.10 ( Fall 2012): 1363-1373 , https://doi.org/10.1057/jors.2011.146

For asymmetric warfare see

Moshe, Kress. "Lanchester models for irregular warfare." Mathematics 8.5 (May 2020), https://doi.org/10.3390/math8050737

[6] Frank E Grubbs and John H. Shuford. "A new formulation of Lanchester combat theory." Operations Research 21.4 (August 1973): 926-941, https://doi.org/10.1287/opre.21.4.926

George A Gamow and Richard E. Zimmerman. Mathematical Models for Ground Combat. Armed Services Technical Information Agency, 1960.

For mathematical models for tank combats, see

Richard H, Peterson. "On the "Logarithmic Law" of Attrition and its Application to Tank Combat." Operations Research 15.3 (June 1967): 557-558 , https://doi.org/10.1287/opre.15.3.557

[7] Harrison C. Schramm and Donald P. Gaver. "Lanchester for cyber: The mixed epidemic-combat model." Naval Research Logistics (NRL) 60.7 (October 2013): 599-605, https://doi.org/10.1002/nav.21555

[8] James G Taylor. "Lanchester-type models of warfare and optimal control." Naval Research Logistics Quarterly 21.1 ( March 1974): 79-106 , https://doi.org/10.1002/nav.3800210107

[9] Pei-Leen, Lie Huai-Ku Sun, and Yue-Tarng You. "A system dynamics model for modern AirLand battle." Asia-Pacific Journal of Operational Research 31.05 (May 2014), https://doi.org/10.1142/S0217595914500316

[10] Moshe, Kress, Jonathan P. Caulkins, Gustav Feichtinger, Dieter Grass, and Andrea Seidl. "Lanchester model for three-way combat." European Journal of Operational Research 264, no. 1 ( January 2018): 46-54, https://doi.org/10.1016/j.ejor.2017.07.026

[11] Stephen. G, Coulson."Lanchester modelling of intelligence in combat." IMA Journal of Management Mathematics 30, no. 2 (January 2019): 149-164, https://doi.org/10.1093/imaman/dpx014

Broadly speaking, there are articles shedding light on the assessment of a model through the lens of PDE(time-space analysis) see

Therese Keane. "Combat modelling with partial differential equations." Applied Mathematical Modelling 35.6 (June 2011): 2723-2735, https://doi.org/10.1016/j.apm.2010.11.057

[12] Andreas J, Novák, , Gustav Feichtinger, and George Leitmann. "On the optimal trade-off between fire power and intelligence in a lanchester model." Dynamic Perspectives on Managerial Decision Making: Essays in Honor of Richard F. Hartl (September 2016): 217-231 https://doi.org/10.1007/978-3-319-39120-5_13 ,

[13] Moshe, Kress and Roberto Szechtman. "Why defeating insurgencies is hard: The effect of intelligence in counterinsurgency operations—A best-case scenario." Operations Research 57, no. 3 (2009): 578-585. https://doi.org/10.1287/opre.1090.0700

[14] Gustav, Feichtinger, Andreas Novak, and Stefan Wrzaczek. "Optimizing counter-terroristic operations in an asymmetric Lanchester model." IFAC Proceedings Volumes 45, no. 25 (September 2012): 27-32, https://doi.org/10.3182/20120913-4-IT-4027.00056

It is worth noting that the attrition rates in the Lanchester model can be transformed into time-dependent parameters, thus making our system a non-autonomous one. For more information see S. G, Coulson, Lanchester

The contribution of our work is to scrutinize the role of different collateral damage functions and their mathematical features, including their convexity, saturation, and non-monotonicity and its effect on the Lanchester model' dynamics. These features all stem from various facets defining a conflict. Paying extra attention to the existing theoretical frameworks offering alternative exegeses will underscore the importance of dealing with sophisticated functions and curves.

The interaction between the Government's ability to target another player and the collateral damage function, echoing the insurgents' overall situation, lies at the heart of our argument, which explains the possibility of multistability in our analysis. We argue that having a different yet comprehensive category of collateral damage, which depends on combat situations, will enrich our understanding of the Lanchester model. The scholarship on the effect of collateral damage typically focuses on its existence and role in the dynamics of the conflict between government and insurgents, thus reemphasizing the role of intelligence in a successful insurgency. Besides, we need to pay extra attention to various theories and conditions according to which collateral damages, emblematic of the functions fueling insurgencies, have different forms, applications, and interpretations. One can exemplify the extent to which these functions can be controlled by governments.[15]. However, as history has laid bare, this control has not been easy or feasible. For instance, the history of the Russian way of counter-insurgency before the collapse of the Soviet Union explained attempts at causing divisions within the targeted society's classes [16] and consequently, its contribution to failure in Afghanistan as a tribal society[17]. Notwithstanding the importance of having a comprehensive category of the geospatial relationship of population and insurgents, we narrow down our focus to deal with cases highlighting the simple mixture of population-insurgents dynamics and the government targeting of this mixture in firing.

In fact, an important theory of counter-insurgency[18], on the basis of historical analysis, sheds light on accommodation and using brute force to control the civilian population offers a different explanation for success in counter-insurgencies in some cases[19]. The government's ability to effectively target its armed rivals is contingent on many issues, ranging from domestic constraints to international issues. [20]. The latter, in addition to international constraints to totally annihilate threats, depends on the extent to which regional equations with time dependent parameters for modelling intelligence collection in adversarial situations." IMA Journal of Management Mathematics (2025), https://doi.org/10.1093/imaman/dpaf012

[15] See Gilbert, Emily. "The gift of war: Cash, counterinsurgency, and 'collateral damage'." *Security Dialogue* 46, no. 5 (2015): 403-421. Note that in the next sections, by discussing Cusp bifurcation, we deal with changes in collateral damage and its impact.

[16] Marshall, Alex. "Counter-Insurgency and the Russian 'Way of War' " in Thomas, M., & Curless, G. (Eds.). (2023). *The Oxford handbook of late colonial insurgencies and counter-insurgencies*. Oxford University Press. p 51

[17] Statiev, Alexander. "Soviet Counter-Insurgency: From the Civil War to Afghanistan In Alexander Hill( ed) *The Routledge Handbook of Soviet and Russian Military Studies*, pp. 433-450. Routledge, p 447

[18] The classification of curves and its mathematical features depends on many factors that is beyond the scope of this article. For seeing different approaches to deal with civilians see Michael G, Findley and Joseph K. Young, "Fighting fire with fire? How (not) to neutralize an insurgency." *Civil Wars* 9, no. 4 (2007): 378-401.

[19] Jaqueline, Hazelton, *Bullets Not Ballots: Success in Counterinsurgency Warfare*. Cornell University Press, 2021, pp 19-22; the idea of isolating and controlling population can be traced in the Tsarist Russia's counter-insurgency policies. See Anthony, Joes, *Urban guerrilla warfare*. University Press of Kentucky, 2007. P 132; For the Russian war on Chechen and its brutality see German, Tracey. "The Chechen Wars." In Alexander Hill( ed) *The Routledge Handbook of Soviet and Russian Military Studies*, pp. 313-324. Routledge, p 321

[20] In some cases, a principal-agent relationship between players will place a premium on some complexities. See Eli, Berman and David A. Lake, eds. *Proxy wars: Suppressing violence through local agents*. Cornell University Press, 2019.

players support insurgents and their ups and downs[21]. Additionally, shifts in insurgencies and the balance of power between rivals can further weaken popular support for an insurgency.[22] This feature reminds us of the importance of studying non-monotonic curves for further assessment in the attempts at better understanding asymmetric conflicts in a dynamic playground and the complicated nature of an insurgency and populace.[23]

One of the most significant results of shedding more light on collateral damage curves is to consider their sensitivity toward damages or the value of C. Attaching importance to a more sophisticated function implies that the effect of small damages is minuscule in fueling more insurgencies in some cases[24]. In a comparison with different forms of curves like linear and quadratic forms, this form, in addition to its slow start, serves an additional benefit; it features saturation in functions, which means that after a threshold $C_T$, the provoking effect of insurgency, in a simplified version, at least remains relatively constant. In fact, in the presence of some geographical and operational constraints, despite benefiting from support from the local population, the curve will be saturated[25] after passing a threshold. A typical form of monotonic damage curve is depicted in Figure 1.

Damages provoke the insurgent agent to react, which is basically to gather around hiring more forces among affected communities and employ the hired forces to extend the conflict. There are various possibilities regarding the qualitative behaviors of such a reaction, as follows:

1) Low magnitude damage does not provoke insurgents until the magnitude passes a threshold, after which the reaction rises sharply.[26] Then after a threshold, due to the small scale of Government involvement and limited sources of recruitment for insurgents, it becomes saturated

2) Even a low magnitude of damage provokes the insurgent to react monotonically, but when damages pass a threshold, the insurgent reaction will reduce due to limited resources, or possible opposition/indifference of affected communities who pay the actual price. See Figure 2 for the general trend of a peak damage curve.

---

[21] John, Mitton, *Why Rivals Intervene: International Security and Civil Conflict*. University of Toronto Press, 2023. P 130; Antonio, Giustozzi, Jihadism in Pakistan Al-Qa'ida, Islamic State and the Local Militants, I.B. TAURIS. 2023, p 41 ; Zafar Iqbal, Yousafzai, Yousafzai, Zafar Iqbal. *The troubled triangle: US-Pakistan relations under the Taliban's shadow*. Routledge India, 2021., pp 99-100; ;Henri J. Barkey, What Role Is Turkey Playing in Syria's Civil War?, Council on Foreign Relations, , December 6, 2024, https://www.cfr.org/expert-brief/what-role-turkey-playing-syrias-civil-war;

[22] For shift in the strategy of the FMLN see Russell, Crandall, *The Salvador Option*. Cambridge University Press, 2016. P 367;

[23] It is important to note that focusing merely on civilian casualties in fueling radicalism will be problematic and requires empirical data clearly indicating this connection. For more information see Aqil, Shah, Do US drone strikes cause blowback? Evidence from Pakistan and beyond. *International Security*, *42*(4), (2018), 47-84, https://doi.org/10.1162/isec_a_00312 . It should be borne in mind that in some complex environments, the inevitable existence of population, in addition to new and accessible technology, can have mix results for operations. See Weissmann, Mikael. "Urban warfare: challenges of military operations on tomorrow's battlefield in Weissmann, Mikael, and Niklas Nilsson (Eds) . *Advanced Land Warfare: Tactics and Operations*. Oxford University Press, 2023, p 147

[24] However, in some circumstances, the initial effect can be high. Nevertheless, saturation is still a possibility.

[25] Saturation can happen as a result of massive destruction, which directly targets insurgents' logistics and leave nothing behind. See Nothing left to bomb': Yemen's civilians bear brunt of US airstrikes on Houthis , The Guardian, May 26, 2025, https://www.theguardian.com/global-development/2025/may/26/yemen-us-israeli-airstrikes-houthi-famine-humanitarian-crisis-civilians ; Boloorani, Ali Darvishi, Mehdi Darvishi, Qihao Weng, and Xiangtong Liu. "Post-war urban damage mapping using InSAR: the case of Mosul City in Iraq." *ISPRS International Journal of Geo-Information* 10, no. 3 (2021): 140., https://doi.org/10.3390/ijgi10030140

[26] On the basis of our forms, we may have sharp beginning

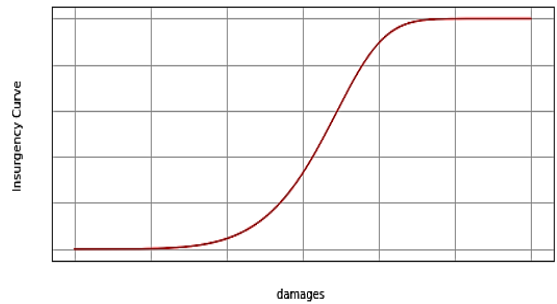

**Figure 1**: A schematic of a sophisticated version of saturated collateral damage. As we are observing, after a specific value of collateral damages, the function becomes constant and its convexity changes. Moreover, the effect of the curve for small values of C is low. In fact, this graph presents an optimistic scenario for insurgents when increased damages cannot go beyond a specific level and yet affect the logistical relationship between insurgents and the population. In this figure, the small or medium scale of the violence prevents the curve from decreasing.

Another type describing collateral damage functions are non-monotonic curves, which emphasizes the level of damage [27] and the behavior of the function[28]. As we laid out earlier, in an attempt to crush insurgencies, given the relatively successful experiences in the contemporary world, brute force as a violent means is one of the pillars of isolating insurgents from the targeted population and sources. Meanwhile, collateral damage and harming the civilian population are an inevitable part of counterinsurgencies. The experience of cruel counter-insurgency measures in different areas, including Guatemala[29] and El Salvador[30] shows the effect of brutality in destroying or controlling the population, due to cutting off the access of the stranded insurgents to the population, and sources [31] could decrease the curve. In the meantime, it can cause the targeted population to turn away from the insurgents.[32] In some examples, it can actuate people and yield more complex results[33]. In the case of the declining effect, notwithstanding facing an increasing curve from

[27] Or brute force

[28] In some cases, by comparing the government-insurgent scenarios for perfect or random shooting, we can see the possibility of diminishing popular support for one side. See Atkinson, Michael P., and Moshe Kress. "On popular response to violence during insurgencies." *Operations research letters* 40, no. 4 (2012): 223-229, p 226, https://doi.org/10.1016/j.orl.2012.04.004

[29] Russell, Crandall, *The Salvador Option*. Cambridge University Press, 2016. P 89;

[30] Jaqueline, Hazelton, *Bullets Not Ballots: Success in Counterinsurgency Warfare*. Cornell University Press, 2021, pp 113-115

[31] And probably support

[32] Plakoudas, Spyridon A., "Mass Killings of Civilians in Counter-Insurgency: Killing More, Winning More?, Infinity Journal, Volume 4, Issue 3, spring 2015, pages 34-38, https://doi.org/10.64148/msm.v4i3.6

To clarify, our models deal with the collateral damage functions as a result of the mixture and vicinity of some non-insurgents to insurgents. Nevertheless, we can trace the sign of targeting civilians without connections to insurgents. See Handy, Jim. "Insurgency and Counter-insurgency in Guatemala." In  Jan L. Flora and Edelberto Torres-Rivas (eds) "*Sociology of "Developing Societies" Central America*", pp. 112-139. London: Macmillan Education UK, 1989, p 122

[33] In the Saudi-led war against Yemen, we can trace the impact of collective punishment campaign. See Shield, R. 2021. "Coercing a Chaos State: The saudi-Led Air War in Yemen." In Air Power in the Age of Primacy: Air Warfare Since the Cold War, edited by Haun, P., C. Jackson, and T. Schultz, 201–228. Cambridge: Cambridge University Press. P 220

the starting point[34], we see that after a threshold, the effect of collateral damages would be on the decline[35]. A typical form of non-monotonic damage curve is depicted in Figure 2. If we denote the damage factor as $C$, then a typical formulation of such curve is $aC^m e^{-kC^n} + b$. Here $a, b, m$ and $k$ are combined system's parameters.

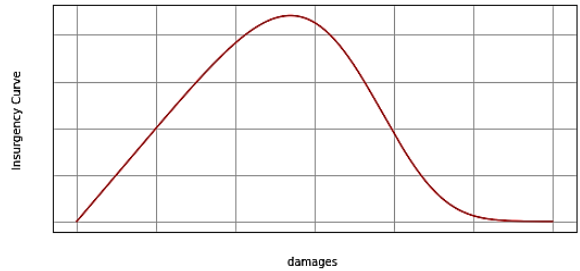

**Figure 2**: A schematic of a sophisticated non-monotonic collateral damage curve according to which, the increase in the value of damage can decrease the provoking effect for the insurgency. In fact, due to the larger scale of G in this curve, an increase in collateral damage function after a peak has resulted in less support for insurgency.

The chart below shows how these parameters change the qualitative behavior of our non-monotonic curve.

| | |
|---|---|
| $a$ | Change the peak of the curve, higher $a$ means higher peak and vice versa |
| $b$ | grievance, innate sympathy toward insurgents |
| $m$ | Affects both peak and initial response. Higher $m$ means lower peak and less sensitivity toward low-level damages and vice versa |
| $k$ | Changes peak, more k means lower peak and vice versa |

In this article, based on reasonable initial conditions and parameters, we conduct simulation to study the behavior of the system.

Within a separate study, real data can be analyzed empirically by evaluating the parameters and detecting the running dynamics. Specifically, the data of the case of approved insurgent control will be very useful

---

We can mention the role of regime type in classifying the curve. However, it is a hotly contested issue. For the effect of autocracy on systematically targeting civilians see Bales, Marius, and Max Mutschler. 2025. "A New Autocratic Way of War? Autocracy, Precision Strike Warfare and Civilian Victimization." *Defence Studies*, June, 1–24. http://dx.doi.org/10.1080/14702436.2025.2531617 ;

[34] It is possible to have a decreasing curve at the outset of conflict. However, the classification of these objects is beyond the scope of this article and its mathematical agenda. For finding this pattern the scholarship on selective and indiscriminate violence in the presence of argument over control and protecting civilians see Kalyvas, Stathis N. *The logic of violence in civil war*. Cambridge University Press, 2006., PP 167 & 204

[35] More interestingly, we can expect a monotonic decreasing curve by committing violence. For the effect of indiscriminate violence on insurgencies see Lyall, Jason. "Does indiscriminate violence incite insurgent attacks? Evidence from Chechnya." *Journal of Conflict Resolution* 53, no. 3 (2009): 331-362, https://doi.org/10.1177/0022002708330881

For the costly success of the Sri Lanka's government against insurgents see Smith, Niel A. "Understanding Sri Lanka's Defeat of the Tamil Tigers." *Joint Force Quarterly: JFQ* 59 (2010): 40.

for a hybrid (theoretical-empirical) analysis. Note that such studies cannot be done solely based on data, and the interpretation of data science analysis requires mathematical modeling.

In the next section, based on the information and points mentioned above, we conduct a stability analysis of a modified version of the Lanchester model and extract the system's complexities and invariant objects by conducting bifurcation analysis. By alluding to the concept of bifurcation, we mean a significant change in the qualitative behavior of the system due to a change in the parameters of the system. The assessment of the change of being significant is done via observing discontinuity of an auxiliary function, which is usually addressed as a singularity. The bifurcation is called local if the focus of the change is a small neighborhood of the state space. When we observe the qualitative behavior beyond small neighborhoods, the occurred bifurcation is called global. In subsequent sections of our two-dimensional system, the considered local bifurcation is Hopf, and the considered global bifurcation is homoclinic.

First, we explain the Hopf bifurcation. Consider a two-dimensional parameterized dynamical system defined by equation $\dot{x} = f(x, \alpha)$ with the state variable $x \in \mathbb{R}^2$, the parameter $\alpha \in \mathbb{R}$, and the vector field $f$, which has an isolated fixed point at $x^*(\alpha)$. Suppose, the Jacobian matrix $\frac{\partial f}{\partial x}(x^*(\alpha))$ has a pair of complex eigenvalues $\lambda(\alpha) = \mu(\alpha) \pm \omega(\alpha)$ such that $\mu(0) = 0$, $\omega(0) > 0$, and $\frac{d}{d\alpha}\mu(\alpha)|_{\alpha=0} \neq 0$. Supposedly, the first Lyapunov coefficient $\ell_1(\alpha)|_{\alpha=0} \neq 0$. Then, the system undergoes a supercritical (*Subcritical*) Hopf bifurcation when $\ell_1(\alpha)|_{\alpha=0} < 0$ ($\ell_1(\alpha)|_{\alpha=0} > 0$): For negative (*positive*) $\alpha$ with small $|\alpha|$, the fixed point $x^*(\alpha)$ is locally asymptotically stable (*unstable*) and for positive (*negative*) $\alpha$ with small $|\alpha|$, the fixed point $x^*(\alpha)$ is unstable (*locally asymptotically stable*) is surrounded with an attracting (*repelling*) limit cycle; see Figure 3 (*Figure 4*) for typical Supercritical (*Subcritical*) Hopf bifurcation diagrams.

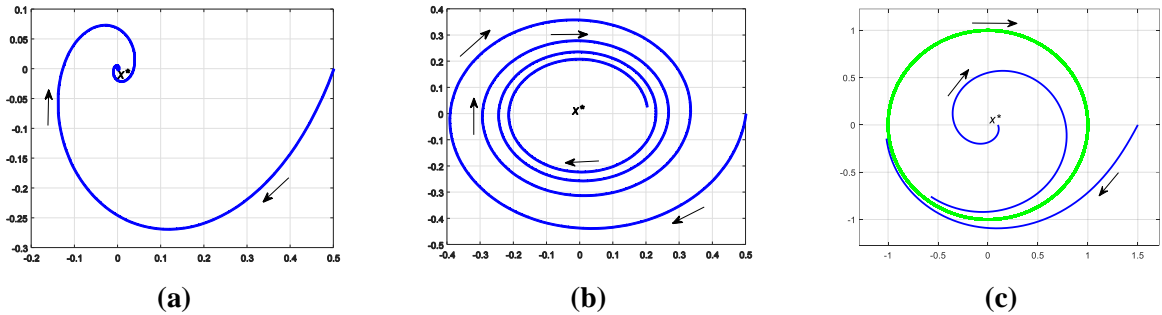

**(a)**    **(b)**    **(c)**

Figure 3: Supercritical Hopf bifurcation (non-catastrophic loss of stability, $\boldsymbol{\ell_1(\alpha)|_{\alpha=0} < 0}$): (a) negative $\boldsymbol{\alpha}$ with small $\boldsymbol{|\alpha|}$. (b) $\boldsymbol{\alpha = 0}$. (c) positive $\boldsymbol{\alpha}$ with small $|\alpha|$.

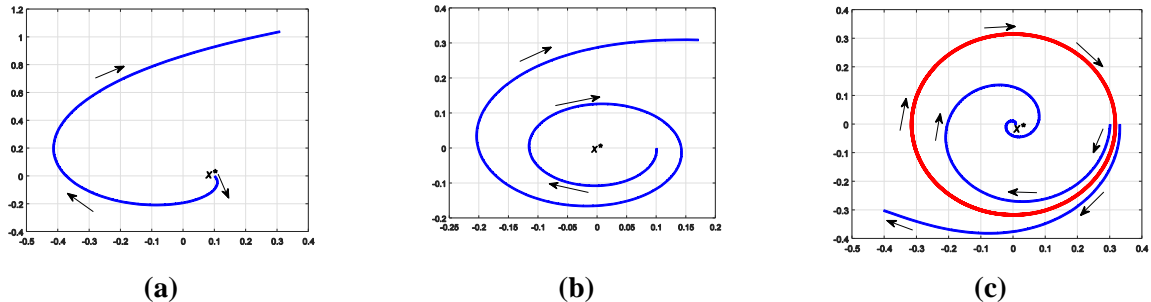

**(a)**    **(b)**    **(c)**

Figure 4: Subcritical Hopf bifurcation (catastrophic loss of stability, $\boldsymbol{\ell_1(\alpha)|_{\alpha=0} > 0}$): (a) negative $\boldsymbol{\alpha}$ with small $\boldsymbol{|\alpha|}$. (b) $\boldsymbol{\alpha = 0}$. (c) positive $\boldsymbol{\alpha}$ with small $|\alpha|$.

Second, we explain the homoclinic bifurcation theorem[36]. Consider a two-dimensional parameterized dynamical system defined by equation $\dot{x} = f(x, \alpha)$ with the state variable $x \in \mathbb{R}^2$, the parameter $\alpha \in \mathbb{R}$, and the vector field $f$ which has an isolated fixed point at $x^*(\alpha)$. Supposedly, the Jacobian matrix $\frac{\partial f}{\partial x}(x^*(0))$ has a real eigenvalues $\lambda_1(0) < 0 < \lambda_2(0)$ such that $x^*(0)$ is a saddle fixed point with a homoclinic connection. Suppose $\sigma_0 = \lambda_1(0) + \lambda_2(0) \neq 0$ with the split function $\beta(\alpha)$. Suppose $\beta'(0) \neq 0$. Then, for all sufficiently small $|\alpha|$, there exists a neighborhood of homoclinic connection in which a unique limit cycle bifurcates from the homoclinic orbit. Then, one of the following cases occurs:

i) The limit cycle exists and is stable for $\beta > 0$ if $\sigma_0 < 0$;

ii) The limit cycle exists and is unstable for $\beta < 0$ if $\sigma_0 > 0$.

See Figure 5 for a typical homoclinic bifurcation diagrams of non-zero saddle quantity

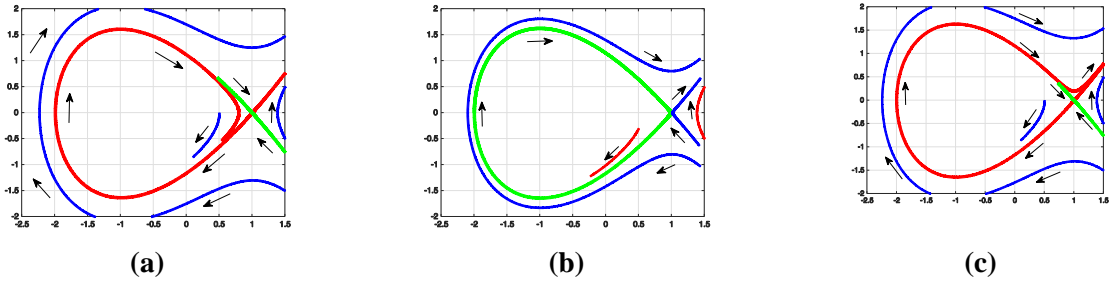

**Figure 5:** Homoclinic bifurcation ($\sigma_0 \neq 0$): (a) $\beta < 0$. (b) $\beta = 0$. (c) $\beta > 0$.

Then, it is crucial to detect the occurrence of homoclionic to address the abovementioned theorem. A powerful method in detecting a homoclinic connection is the method of Melnikov, which is based on perturbation of Hamiltonian dynamical systems. A $2n-$dimential Hamiltonian dynamical system is defined by a set of differential equations $\left( \dot{p} = \frac{\partial H}{\partial q}, \ \dot{q} = -\frac{\partial H}{\partial p} \right)$, where $(p, q) \in \mathbb{R}^n \times \mathbb{R}^n$ is the sate variable and $H: \mathbb{R}^n \times \mathbb{R}^n \to \mathbb{R}$ is the Hamiltonian function of the system. The equations of the system is compactly written as $\dot{x} = JH_x$. Here $x = (p, q)$, $J = \begin{pmatrix} 0 & I_n \\ -I_n & 0 \end{pmatrix}$, $H_x = \frac{\partial H}{\partial x}$

, and $I_n$ is the identity operator of $\mathbb{R}^n$. Suppose that the underlying system is $\dot{x} = JH_x + \epsilon g(x, t)$ for a small enough $\epsilon$. Suppose for $\epsilon = 0$, the unperturbed system possesses a homoclinic orbit to a hyperbolic saddle fixed point $x^*$. Suppose the interior of the region $\Omega$ with $\partial \Omega = \{x^0(t): t \in \mathbb{R}\} \cup \{x^*\}$ is filled with a continuous family of periodic orbits $x^\alpha(t)$ with period $T_\alpha$ for $\alpha \in (-1, 0)$ such that with $h_\alpha = H(x^\alpha(t))$ we have $\frac{d}{dh_\alpha} T_\alpha > 0$. Also, suppose $\lim\limits_{\alpha \to 0} dist(x^\alpha, \partial \Omega) = 0$ and $\lim\limits_{x \to -1} x^\alpha = x_*$ where $x_*$ is a non hyperbolic non-attracting Lyapunov stable fixed point; see Figure 6. Suppose we move $\epsilon$ away from zero, then the new hyperbolic saddle fixed point $x_\epsilon^*$ appears near the previous fixed point $x^*$ of the unperturbed system with non-intersecting stable and unstable manifolds; see Figure 7. It can be proved that the distance between stable and unstable manifolds of $x_\epsilon^*$ at the Poincaré section $t = t_0$ satisfies $d(t_0) = \frac{M(t_0)}{|JH_x(x^0(0))|}$ where

[36] Kuznetsov, Yuri A. *Elements of applied bifurcation theory*. New York, NY: Springer New York, 2004, Theorem 6.1, p 200.

$$M(t_0) = \int_{-\infty}^{+\infty} J H_x(x^0(t-t_0)) \wedge g(x^0(t-t_0), t) \, dt.$$

Here we have Theorem 4.5.3,[37] which states that if $M(t_0)$ has a simple zero and is independent from $\epsilon$ for $\epsilon > 0$ sufficiently small, then the stable and unstable manifolds transversally intersect each other and form a hommoclinic connection to the hyperbolic saddle fixed point $x_\epsilon^*$. See Figure 7 for a typical illustration of the distance function of the Poincaré cross-section.

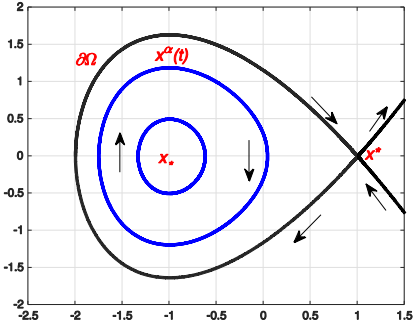

**Figure 6:** The unperturbed Hamiltonian system possessing a Homoclinic connection to the saddle fixed point $x^*$.

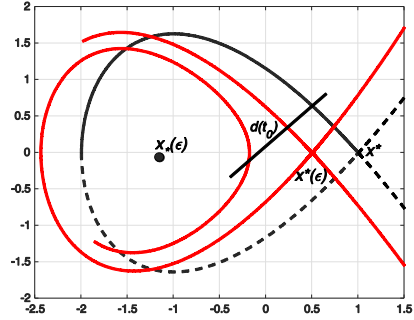

**Figure 7:** For the perturbed system, the distance function between stable and unstable manifolds of the hyperbolic saddle fixed point for $\epsilon \neq 0$.

## Modeling

Let $\mathbb{G}$ and $\mathbb{I}$ be two players, namely government and insurgents. Suppose the variable $G = [\mathbb{G}]$ the number of government combatants. For insurgents, suppose the variable $I = \frac{[\mathbb{I}]}{[\mathbb{P}]}$ represents the portion of the targeted population $\mathbb{P}$ represented in insurgents. Both variables $G$ and $I$ change over time. Here, $[\mathbb{G}]$, $[\mathbb{I}]$, and $[\mathbb{P}]$ denote the quantity of populations $\mathbb{G}$, $\mathbb{I}$, and $\mathbb{P}$ respectively. Let's denote the collateral cost by $C$. We employ the following four sets of assumptions. The cost of the presence of $\mathbb{I}$ and its conflict with $\mathbb{G}$ is denoted by $\mathbb{C}$ with quantity $C = [\mathbb{C}]$.

In a deviation from the traditional understanding of the mathematical formation of the collateral damage (C), which represented the probability of targeting a civilian randomly in the absence of intelligence, and in tandem with the type of our collateral damage functions that draw on the mixture of some portion of the local population and insurgents. While the former resulted in the decreasing effect of I on the level of C, the latter shows a more complex feature. In fact, by applying the rationale used to study epidemiological systems and the effect of the ill population on suspicion as $\omega SI$ in a given community[38]. For the sake of simplicity, we assume that insurgents are among or in contact with the civil population, but not all civilians are in contact with insurgents. Therefore, with $\mu$, the government can pinpoint insurgents and target them. On the other hand, under the circumstance we set out, we may target insurgents with $(1 - \mu)$, but it will

[37] Guckenheimer, John, and Philip Holmes. *Nonlinear oscillations, dynamical systems, and bifurcations of vector fields*, Springer, 2013. P 188.

[38] For epidemiological models and the dynamics of suspicious, ill, and recovered individuals see Brauer, Fred, Carlos Castillo-Chavez, and Zhilan Feng. *Mathematical models in epidemiology*. Vol. 32. New York: Springer, 2019.
In the meantime, there is a trend of modeling conflicts through the prism of epidemic understanding of government and insurgent's violence and the problem of civilian choice as an epidemic model see Zhukov, Yuri M. "An epidemic model of violence and public support in civil war." *Conflict Management and Peace Science* 30, no. 1 (2013): 24-52.;

come at a cost, which refers to targeting civilian population at a rate of $I(1-I)$[39]. Therefore, we have these axioms

A1: The zero level insurgent, which is $I = 0$, means no[40] collateral cost in the absence of insurgents and logistical prerequisites ; namely, $\boldsymbol{C} = 0$[41].

A2: When $I = 1$, there would be no room for civilian damage, thus no cost will be sustained; i.e., $\boldsymbol{C} = 0$.

A3: The cost value, i.e., the damage factor, is proportional to the size of government troops and the infusion of insurgents; that is $\boldsymbol{C} \propto [(1-I)I]G$. The equator of the proportional relation is the quantity of $\nu$ denoted by $\nu = 1 - \mu$ in which $\mu$ is represents the knowledge of $\mathbb{G}$ about $\mathbb{I}$. Thus, the cost value becomes $\boldsymbol{C} = (1-\mu)[(1-I)I]G$. Meanwhile, starting from $I = 0$, an increase in the value of I among the local population will result in an increment in $\boldsymbol{C}$. After $I = 0.5$, the value decreases. This gives rise to complexity. We differentiate between the attrition rate of targeting insurgent combatants and that of collateral damage curves, which deals with a relatively different phenomenon[42]

A4: We assume that the insurgent body $\mathbb{I}$ reacts to the cost via the reaction functional $\theta$, as a function of the cost powered by the intensity factor $\gamma_2$; namely, $C = \gamma_2 \boldsymbol{C} = \gamma_2(1-\mu)[(1-I)I]G$. The functional is formed by a combat-curve-type of Figure 2. Therefore, $\theta = \theta(C) =$

$\theta(\gamma_2(1-\mu)[(1-I)I]G)$. Note that $\gamma_2$ may be taken as a function of $\gamma_1$, but due to its very slow variation, for simplicity, $\gamma_2$ is considered as a constant. As we mentioned earlier, we can have b[43] as a grievance or ideological factors further fueling an insurgency. Depending on its structure and values, it can have a far-reaching impact on the tensions between the government and insurgents. Note that the typology of collateral damage functions has an embedded ideology that explains its variation among different insurgent groups. Therefore, when it comes to assessing ideology or grievance in modeling, we need to mention a general form. Basically, $\theta(0) \neq 0$ indicates a constant value of grievances of the targeted area toward the government as the adversary. However, deeming ideology as a main driver of insurgency and overlooking desperation and instability is not true[44]. In fact, the presence of grievances can keep tensions and revenge aflame.[45] In a similar vein to the general type of the Lanchester model, with the collateral damage as[46]

$$\begin{cases} \dot{G} = -\alpha I + \beta \\ \dot{I} = -\gamma_1 G(\mu + (1-\mu)I) + \eta\theta(C) \end{cases} \tag{1}$$

---

[39] The scholarship on epidemic modeling of the Lanchester model of warfare gives credence to the mixture of non-insurgents and insurgents as a new drive for insurgency. However, its main focus is recruitment not targeting insurgents in the presence of population. For the epidemic effect of insurgents' recruitment of population see Howell, Jeffrey M. "Modeling insurgency attrition and population influence in Irregular Warfare." PhD diss., Monterey, California. Naval Postgraduate School, 2007.

[40] or realistically, miniscule effect

[41] As we show later, I does not mean that it has no effect on fueling insurgency.

[42] Even though we want to remain committed to the generally accepted format of C, the existence of $\eta$ as a parameter of the $\theta$ will help us conduct our bifurcation analysis and study the system's global bifurcation. It is worth mentioning that targeting civilians and controlling the population in some cases is weakly related to gathering information, specifically in cases where the sensitivity toward the population is not high.

[43] We can have b $= 0$ or $b \neq 0$

By a time re-parameterization, it is possible to set $\eta = 1$.

*A remark on modeling*: Population dynamics modeling is common in biology and ecology. In Biological population dynamics, there is a firm principle that when there is no population of one kind (the zero population), then there will be no population of that kind along the forward time. This principle applies to the right-hand side of the model to ensure that the positive region remains invariant dynamically. Furthermore, in population dynamics modeling, the model may be restricted to perform some boundedness properties. Like the line of Lanchester modeling, here, we do not apply such principles; thus, the presented model in equation (1) is not assumed to comply with the property of invariant positive region nor the boundedness restriction. However, when the values of $G$ or $I$ go beyond it means that the dynamics go beyond the Lanchester-based interpretation and would require a new line of modeling. In practice, there is no universal modeling for a single real-world phenomenon. As an example, in physics, many presented models are valid within a specific scale of validity, such as classical mechanics, quantum mechanics, or special/general relativity. It is also worth noting that, in Lanchester modeling, the general properties of involved functions may have various realizations. Different forms of realization of a specific functionality may be used for the means of simulation.[47]

Now, we conduct the equilibrium and stability analysis for system (1). The equilibria are the solutions of $\dot{G} = 0$ and $\dot{I} = 0$ which leads to the following system of algebraic equations

---

[44] See Brahimi, Alia. "Ideology and terrorism." In *Chenoweth, Erica, Richard English, Andreas Gofas, and Stathis N. Kalyvas, eds. The Oxford handbook of terrorism. Oxford University Press, 2019, p 311, https://doi.org/10.1093/oxfordhb/9780198732914.013.17*

[45] In the Richardson model of arms races, we have a constant named grievance that explains the constant amount of players' defense expenditure in the absence of provocations. See Richardson, L. F. (1960). *Arms and Insecurity: A Mathematical Study of the Causes and Origins of War*. London: Stevens & Sons; Pittsburgh, PA: Boxwood Press; Chicago, IL: Quadrangle Books ; [45] In some cases, despite the success in suppressing insurgency, which gives rise to the saturation rate, the emergence of new threats is possible due to the existence of a potentially vengeful population. See Hashim, Ahmed. *When counterinsurgency wins: Sri Lanka's defeat of the Tamil Tigers*. University of Pennsylvania Press, 2013, p 79

[46] The general model of counterinsurgency deals with the zero ideology and decreasing nature of C as I increases. However, our model has a different stance on C, stating that increase in I could increase G at the outset of insurgent's recruitments. For the general form of Lanchester model with CD function see Kress, M., & Szechtman, R. (2009). Why defeating insurgencies is hard: The effect of intelligence in counterinsurgency operations—A best-case scenario. *Operations Research*, *57*(3), 578-585, p 580, https://doi.org/10.1287/opre.1090.0700

[47] For positive invariance in biological models see Murray, James D. *Mathematical biology: I. An introduction*. Vol. 17. Springer Science & Business Media, 2007.

$$
\begin{cases}
I = \dfrac{\beta}{\alpha} \\[2mm]
\gamma_1 \left( \mu + (1-\mu)\left(\dfrac{\beta}{\alpha}\right) \right) G = \theta \left( \gamma_2 (1-\mu)\left(\dfrac{\beta}{\alpha}\right)\left(1 - \dfrac{\beta}{\alpha}\right) G \right) \\[2mm]
\text{-----------------------} \\[1mm]
I^* = \dfrac{\beta}{\alpha}, \quad G^* = G^*(\alpha, \beta, \mu, \gamma_1, \gamma_2) \subset \mathbb{R} \setminus \{0\}
\end{cases}
\tag{3}
$$

Finding the values of $G^*$ is contingent on solving of an equation displaying the intersection of a quadratic and a sophisticated non-linear function of $G^*$. In order to find the equilibria of this system and take into consideration the possibility of the occurrence of multistability and, consequently, global bifurcations, we need to have a complete category of this intersection. Let $\gamma_1$ be the parameter for changing the linear function in order to study the different forms of the intersection. By decreasing the value of $\gamma_1$, we will have different categories with various equilibria. The typical form of equilibria for a non-monotone damage curve is illustrated in Figure

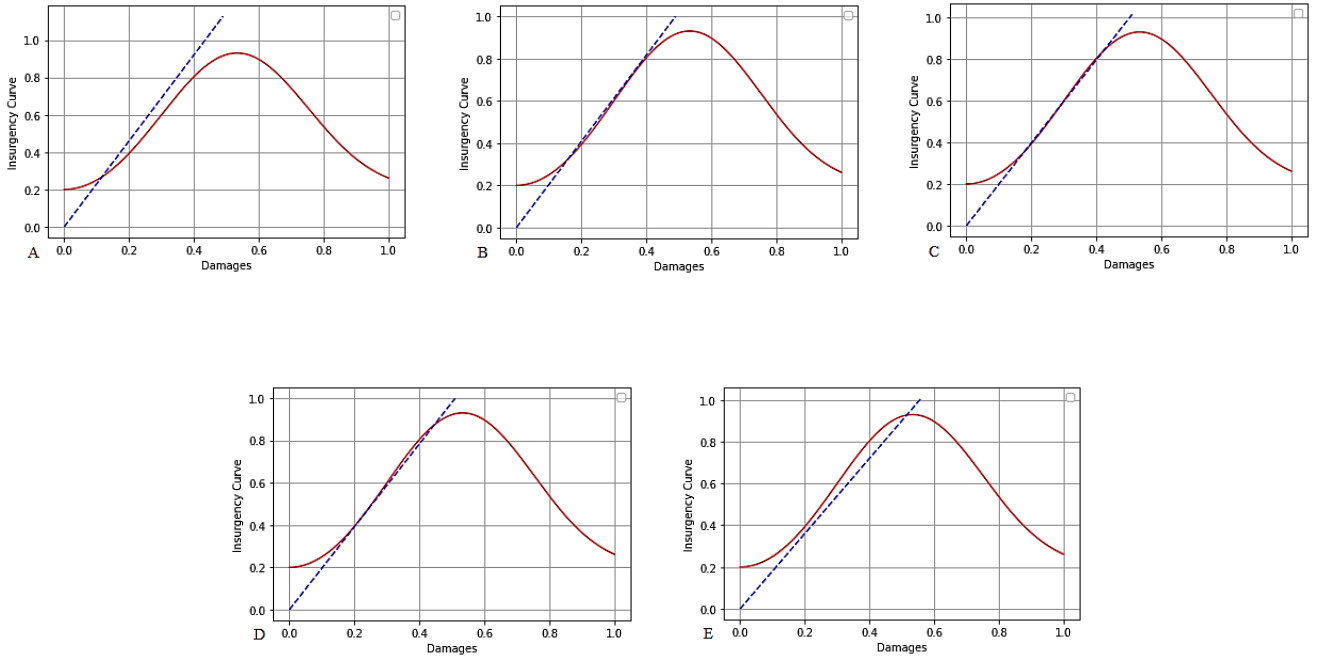

**Figure 8:** The category of the intersection of linear and nonlinear functions as a result of change in the value of $\gamma_1$

For $\gamma_1 > \gamma_1^0$, Figure 3(A), we have one equilibrium point. For $\gamma_1 = \gamma_1^0$, Figure 3(B), we have two equilibria. The interval $\gamma_1^0 < \gamma_1 < \gamma_1^1$, Figure 3(C), gives us three equilibrium points. Like Figure 3(B), the equality $\gamma_1 = \gamma_1^1$, in Figure 3(D), generates two equilibrium points and, finally, $\gamma_1 < \gamma_1^1$, Figure 3(E), gives us only one equilibrium point. Therefore, $G^*$ could be a singleton, doubleton, or tripleton. Having discussed different possibilities of the number of equilibria in our system by reducing the value of $\gamma_1$[48], we now conduct the stability analysis of our system to assess the stability of our equilibrium points and their geometric positions. By defining the Jacobian matrix, we will have

$$J = \begin{bmatrix} \mathbf{0} & -\alpha \\ -\gamma_1(\mu + (1-\mu)I^*) + \left(\dfrac{\partial\theta}{\partial C}\dfrac{\partial C}{\partial G}\right)_{I^*,G^*} & -\gamma_1(1-\mu)G^* + \left(\dfrac{\partial\theta}{\partial C}\dfrac{\partial C}{\partial I}\right)_{I^*,G^*} \end{bmatrix} \tag{4}$$

$$= \begin{bmatrix} \mathbf{0} & -\gamma_1(\mu + (1-\mu)I^*) + \left(\dfrac{\partial\theta}{\partial C}\right)_{I^*,G^*}(\gamma_2(1-\mu)[(1-I^*)I^*]) \\ -\alpha & -\gamma_1(1-\mu)G^* + \left(\dfrac{\partial\theta}{\partial C}\right)_{I^*,G^*}(\gamma_2(1-\mu)[1-2I^*]G^*) \end{bmatrix}^T$$

$$\tau = \mathbf{Trace}(J), \delta = \mathbf{Det}(J)$$

---

[48] We could use $\theta$ as our goal parameter in assessing the system's equilibria. Contrary to the case of $\gamma_1$, an increase in the value of $\theta$, created this order.

The characteristic polynomial equation, $F(\lambda) = \lambda^2 - \tau\lambda + \delta = 0$ determines the possible local stability. Each equilibrium needs to be separately analyzed. Here, we are specifically interested in the stability of equilibria when $\gamma_1^0 < \gamma_1 < \gamma_1^1$ ; see Figure 3 (C). We gave $G_1^* < G_2^* < G_3^*$ and three equilibria $\left(I^* = \frac{\beta}{\alpha}, G_i^*\right)_{i=1,2,3}$. The equilibria $(I^*, G_1^*)$ and $(I^*, G_3^*)$ are saddle. Only the equilibria $(I^*, G_2^*)$ is the point of interest for bifurcation. The Hopf bifurcation occurs when $\tau = 0$ while $\delta > 0$; that is:

$$\left(\frac{\partial\theta}{\partial C}\right)_{I^*,G^*} = \frac{\gamma_1(1-\mu)G^*}{(\gamma_2(1-\mu)[1-2I^*]G^*)} \tag{5}$$

$$\frac{\beta}{\alpha} < \frac{-\mu + \sqrt{\mu}}{1-\mu} \tag{6}$$

Therefore, equations (4)-(5) give the necessary condition of Hopf bifurcation for the equilibrium $(I^*, G_2^*)$. The sufficient condition depends on the value of the first Lyapunov coefficient. A typical multiple-fold bifurcation caused by the non-monotonic damage function is depicted in Figure 4.

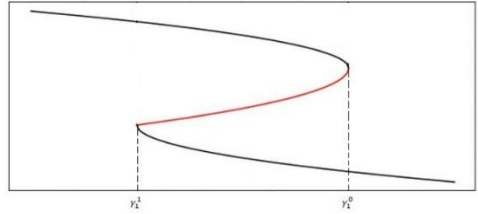

**Figure 9**: The occurrence of Fold bifurcation (at $\gamma_1^0$ and $\gamma_1^1$) in our system as $\gamma_1$ decreases. We can observe the possibility of a Hopf bifurcation in the red area, where there are three intersections.

As illustrated in Figure 4, we can conclude that the Fold bifurcation occurs at two points, and the occurrence of three intersections, which is a prerequisite for the Hopf bifurcation. The red area shows the possibility of the Hopf bifurcation if (5) and (6) are met. To compute the values of $\gamma_1^1$ and $\gamma_1^0$, at first we solve the equation $C \theta'(C) - \theta(C) = 0$ which has two positive solutions $0 < C_1 < C_0$. Then,

$$\gamma_1^1 = \left(\frac{(1-\mu)\frac{\beta}{\alpha}\left(1-\frac{\beta}{\alpha}\right)}{\mu + (1-\mu)\frac{\beta}{\alpha}}\right)\gamma_2\theta'(C_1), \qquad \gamma_1^0 = \left(\frac{(1-\mu)\frac{\beta}{\alpha}\left(1-\frac{\beta}{\alpha}\right)}{\mu + (1-\mu)\frac{\beta}{\alpha}}\right)\gamma_2\theta'(C_0) \tag{7}$$

are achieved by some algebras. Here, the Hopf bifurcation appears at $\gamma_1^H \in (\gamma_1^1, \gamma_1^0)$. For the case of the given typical exponential damage function $\theta(C) = aC^m e^{-kC^n} + b$ we take the set of values $a = 5, m = 2, k = 4, n = 3$, and $b = 0.2$ for the means of simulations. Based on this form of $\theta$ and some algebra, we will have $C_1 \approx 0.218278$, $C_0 \approx 0.379127$. Then, for a given values $\gamma_2 = 8.33, \alpha = 0.95, \beta = 0.4, \mu = 0.5$, we will have, $p = 243675/348194$, and $\gamma_1^1 \approx 2.80525$, $\gamma_1^0 \approx 2.93208$. See Figure 5 for various $\gamma_1$. In Figure 5, (a) shows the phaseportrait for $\gamma_1 < \gamma_1^1$, (b) shows the phaseporteaite for $\gamma_1^1 < \gamma_1 < \gamma_1^H$ after supercritical Hopf Bifurcation and the occurrence of the limit cycle, (c) shows the phase portrait for $\gamma_1 > \gamma_1^0$. Disappearance of the limit cycle happens by increasing the parameter $\gamma_1$ within the region $(\gamma_1^1, \gamma_1^0)$.

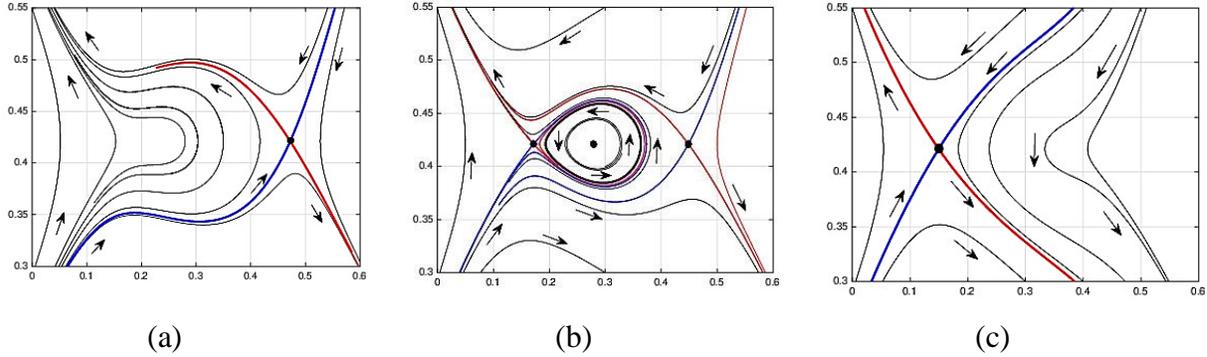

(a)              (b)              (c)

**Figure 10** : The simulation for system of equation (1) with $\theta(C) = \left(5C^2\, e^{-4C^3}\right) + 0.2$, and parameters , $\boldsymbol{\gamma_2 = 8.33}$ , $\boldsymbol{\alpha = 0.95}$, $\boldsymbol{\beta = 0.4}$, $\boldsymbol{\mu = 0.5}$. (a) $\boldsymbol{\gamma_1 = 2.80 < \gamma_1^1}$, (b) $\boldsymbol{\gamma_1 = 2.8571 \in (\gamma_1^1, \gamma_1^1)}$. (c) $\boldsymbol{\gamma_1 = 2.95 > \gamma_1^0}$.

Having a periodic $(G, I)$ −dynamics, in Figure 5(b), shows that in the fight, we can have different phases, namely, increasing terror phase, dominant terror phase, recovery phase, and dominant government phase[49], which explain the dynamics of the battle between the two adversaries. More importantly, the dynamics of the conflict, given the non-monotonic nature of the curve, suggest the pitfalls of falling into the risky zone of the three equilibria state, according to which a periodic situation between two forces, with the change of parameters as a fact, can bring about collisions and end up in extinction. Thus, the question is in the presence of two saddle nodes surrounding the system's limit cycle, how a collision can terminate each of these phases, thus ending up in severe outcomes[50]. In many cases, where players have fallen into the trap of periodic solutions, which is rampant in various conflicts, changes in some parameters, including recruitment rate, the capability to effectively target $I$, or attempts at controlling and isolating local populations, can change the limit cycle and consequently, yield complexities like homoclinic bifurcation.

In this case, we use $\alpha$ as the targeted parameter to show an increase in its value can result in a homoclinic bifurcation. In fact, such an increase signals insurgents' success in targeting the enemy's forces. In the example we provided, increase in $\alpha$ can push the expanding limit cycle westward, while in another situation, given the difference in $\gamma_1$ or the exponentials base, increase in $\alpha$ can push the cycle westward and bring about different ramifications. In fact, due to the existence of saddle nodes in the vicinity of the cycle, regardless of its phase, it can break the cycle and bring the system's trajectories somewhere else.

---

[49] Novák, Andreas J., Gustav Feichtinger, and George Leitmann. "On the optimal trade-off between fire power and intelligence in a lanchester model in Dawid, Herbert, Karl F. Doerner, Gustav Feichtinger, Peter Kort, and Andrea Seidl(Eds) . *Dynamic perspectives on managerial decision making*. Berlin: Springer, 2016. PP 228-229,

[50] In the case of Afghanistan, the selection-destruction cycle, in concurrence with dwindling effectiveness and international support, and abuses that accelerated the collateral damage curve, resulted in the 2021 collapse. See Kilcullen, David. "The Case of Afghanistan: How Wars End 1." In *Kingsbury, Damien, and Richard Iron, eds. How wars end: Theory and practice. Routledge*, 2022. PP 125-128

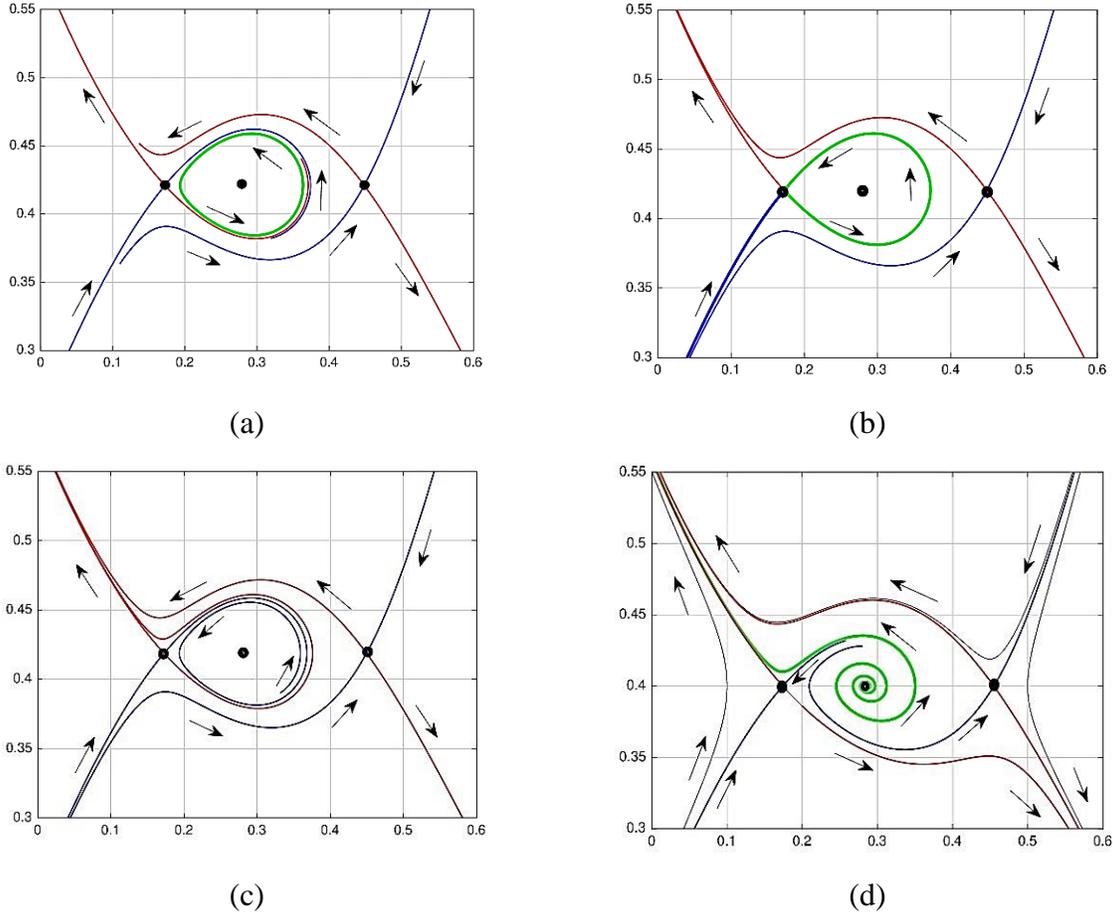

(a)   (b)   (c)   (d)

**Figure 11**: From A to D, we can observe the effect of increase in $\alpha$ on the system's limit cycle in the presence of a saddle node. In D, we our system is undergoing homoclinic bifurcation. In this system we have $\gamma_1 = 2.8571 \in (\gamma_1^1, \gamma_1^0)$, $\gamma_2 = 8.33$, $\beta = 0.4$, $\mu = 0.5$. (a) $\alpha = 0.95 < \alpha_H$, (b) $\alpha = 0.95136 \approx \alpha_H$, (c) $\alpha = 0.9550 > \alpha_H$, (d) $\alpha = 1$.

The mechanism of victory and defeat is asymmetric in $G - I$ dynamics. For $G$, victory has a 2-possibilties. The first is $I(t) \to 0$ as $t$ increases. Practically, this means that for predefined $\varepsilon_0$, the relation $\left\|I(t)\right\| < \varepsilon_0$ is achieved in a plausible time window $[0, \tau]$ and to be guaranteed for arbitrary large $t$. For example, a low initial condition for I, or the collision of the limit cycle with the equilibrium located at the right, yields victory for G. The second is to force insurgent into an acceptable confined region of activity; namely, providing a bounded acceptable region $\Omega$, within which the variable $I$ stays. That is $I(t) \in \Omega$ for all $t > 0$. The government may attempt to reschedule the dynamics and reconstruct the confined region through time, such that $\Omega$ depends on time so that $\Omega_t \to \Omega_0$ as $t$ increase with the property that within $\Omega_0$ the relation $\left\|I(t)\right\| < \varepsilon_0$ is achieved. Both of these strategies have their special enforcing conditions. Conversely, for $I$, victory has the meaning of not-get-confined as well as achieving $\left\|I(t)\right\| > M$ for $t$ less than a limited-defined time window, let say for $t < \tau^*$.

As we saw in Figure 6, in a Lanchester model of warfare with sophisticated $\theta$, even an Increase in the effectiveness of targeting $G$, under the circumstances we laid out above, will lead us to the occurrence of a Homoclinic bifurcation and even undesired repercussions provided that the initial conditions for both variables is trapped into the cycle . In (a), we have a limit cycle that explains the 4 stages mentioned in the scholarship. In (b) Improving the effectiveness of targeting $G$ by increasing $\alpha$, will expand the cycle and cause the global bifurcation. In (c) and (d), we are seeing the limit cycle and periodic solution between the two players fades and the loss of one player[51] will happen.[52] For instance, if G forces the system's initial conditions to fall into the cycle before the beginning of the battle, a closer distance between the second and the third equilibrium($G_2 < G_3$), increase in $\alpha$ can result in the loss of I, while colliding with the first equilibrium , as we saw in figure6, can yield the victory of I.

More importantly, the shape of the curve, as the activator of I, has an important effect on the loss/victory of insurgents. As we showed in figure 2, too much violence can diminish support for I, thus facilitating its loss. Simply arguing, if $\gamma_1 G(\mu + (1 - \mu)I) > \theta(C)$, then we have $\dot{I} < 0$. Admittedly, a decreasing curve will help $G$ reach victory. The caveat here is that some elements, like pressure on imposing $\gamma_1$, the scale of government's involvement in the conflict, and the high value of the curve's peak, may prevent G from exerting excessive violence. On the other hand, the curve's values can turn the screw on G ( $\dot{G} < 0$ or $\beta < \alpha I$ ). In these situations, like the emergence of the Homoclinic bifurcation at the left, we can expect the loss of $G$.

Now, we need some analytic methodology to fully grasp the occurrence of homoclinic bifurcation in our system. To this end, we need to conduct the Melnikov function analysis of our model to study global bifurcation. We use the Melnikov method to detect a homoclinic bifurcation when $\gamma_1^0 < \gamma_1 < \gamma_1^1$;

Since our model is not Hamiltonian, we need to introduce the model as a perturbation of a Hamiltonian dynamical system[53] that we can calculate the Melnikov function and spot the Homoclinic bifurcation in our dynamical systems. A careful look into equation (1) will enable us to conclude that attempts at finding the value of $G^*$ in $\dot{I}$, based on the structure of the model, can resemble that of an unperturbed Hamiltonian system if we use

$$H(G, I) := -\frac{1}{2}\alpha I^2 + \beta I + \frac{1}{2}\gamma_1 \mu G^2 - \int \theta\big(C(G, I^*)\big)\, dG \tag{8}$$

The related Hamiltonian system is

$$\dot{G} = \frac{\partial H}{\partial I} = -\alpha I + \beta \quad , \quad \dot{I} = -\frac{\partial H}{\partial G} = -\gamma_1 \mu G + \theta\big(C(G, I^*)\big) \tag{9}$$

Notwithstanding sharing characteristics, like equilibria, with our model, there is a nuanced difference between our original model and the Hamiltonian version of the Lanchester model of warfare with sophisticated collateral damages. In fact, in this ideal case, we assume that having a good understanding of the targeted environment[54], given the high level of intelligence, will necessitate players to tolerate some levels of collateral damage, which is central to controlling the targeted population. Meanwhile, perturbation in this system stems from the obscured part of the interaction between $G$ and I.

---

[51] In this case $G$

[52] Counterintuitively, if the value of exponential base was different and the middle equilibrium was close to the right equilibrium, increase in $\alpha$ would facilitate the extinction of $I$. In fact, this change in the parameters of the collateral damage curve signal a bifurcation that goes beyond the scope of this article

[53] For more information see Kuznetsov, Yuri A. *Elements of applied bifurcation theory*. New York, NY: Springer New York, 2004, p 231

[54] In this case the level of intelligence($\mu$), is relatively high

Note that based on the mean value theorem, for some $\in [0,1]$ , where $I_\zeta = \zeta I + (1-\zeta)I^*$ we have

$$\theta\big(C(G,I)\big) - \theta\big(C(G,I^*)\big) = \left(\frac{\partial\theta}{\partial C}\frac{\partial C}{\partial I}\right)_{G,I_\zeta}(I-I^*) = \left(\left(\frac{\partial\theta}{\partial C}\right)_{G,I_\zeta}\right)\big(\gamma_2(1-\mu)(1-2I_\zeta)\big)(I-I^*) \tag{10}$$

Here, $\zeta$ is a function of $I$ and $I^*$. Then, we assume that $\mu$ is large enough such that $0 \ll \mu < 1$ so that $0 < (1-\mu) \ll 1$. Based on $H(G,I)$, we can rewrite our original model as a perturbation of a Hamiltonian dynamical system.

$$\begin{bmatrix}\dot{G}\\\dot{I}\end{bmatrix} = \overbrace{\begin{bmatrix}\dfrac{\partial H}{\partial I}\\[6pt]-\dfrac{\partial H}{\partial G}\end{bmatrix}}^{f(G,I)} + \frac{\varepsilon}{(1-\mu)}\overbrace{\left(-\gamma_1\begin{bmatrix}\mathbf{0}\\G\,I\end{bmatrix} + \left(\left(\frac{\partial\theta}{\partial C}\right)_{G,I_\zeta}\right)\big(\gamma_2(1-2I_\zeta)\big)(I-I^*)\begin{bmatrix}\mathbf{0}\\\mathbf{1}\end{bmatrix}\right)}^{g(G,I)} \tag{11}$$

Let us denote $z = (G, I)$. We show that the Hamiltonian system $\dot{z} = f(z)$ has a Homoclinic connection. We denote $\omega = \gamma_2(1-\mu)\left[\left(1-\frac{\beta}{\alpha}\right)\frac{\beta}{\alpha}\right]$. Then, the equilibrium problem of the Hamiltonian system is $f(z) = 0$ and has the solution $(G^*, I^*)_h$ with $I_h^* = \frac{\beta}{\alpha}$ and $\theta(\omega\ G_h^*) = \gamma_1\mu\ G_h^*$. The latter coincides with the case in Figure 3 (B & D) and has two distinct solutions $G_p^*$ and $G_q^*$. Therefore, we have two equilibria $(G_p^*, I_h^*)$ and $(G_q^*, I_h^*)$. Suppose $G_p^* < G_q^*$; see Figure 7. The solution of the homoclinic connection satisfies the relation $H(G,I) = H\big(G_q^*, I_h^*\big)$. This defines $z_h(t) = \big(G_q(t), I_h(t)\big)$.

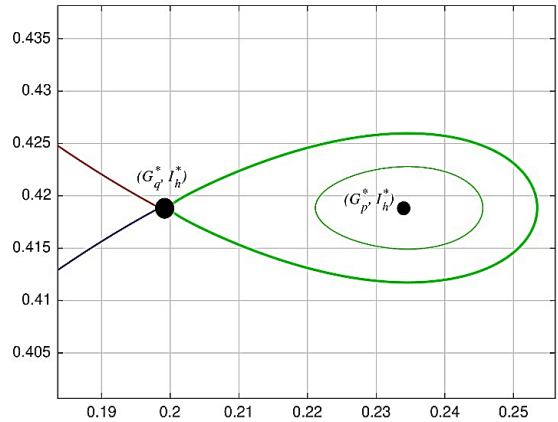

**Figure 12:** The Homoclinic connection of the general Hamiltonian system $\dot{z} = f(z)$. We can see our Hamiltonian system resembles the original system

Now we conduct the Melnikov function analysis. For our unperturbed system, we have the general form of $M(t_0)$ as follows [55].

$$M(t_0) = \int_{-\infty}^{+\infty} \nabla H\big(z_h(t)\big).\,g\big(z_h(t)\big)\,dt \tag{12}$$

[55] Guckenheimer, John, and Philip Holmes. *Nonlinear oscillations, dynamical systems, and bifurcations of vector fields*, Springer, 2013. P 190

Given the complex nature of the model, we should assess this model numerically to calculate the Melnikov function. This integral calculates the distance between the stable and unstable manifolds of the targeted saddle point. Upon the collision, which zeros the distance, a homoclinic bifurcation will appear.

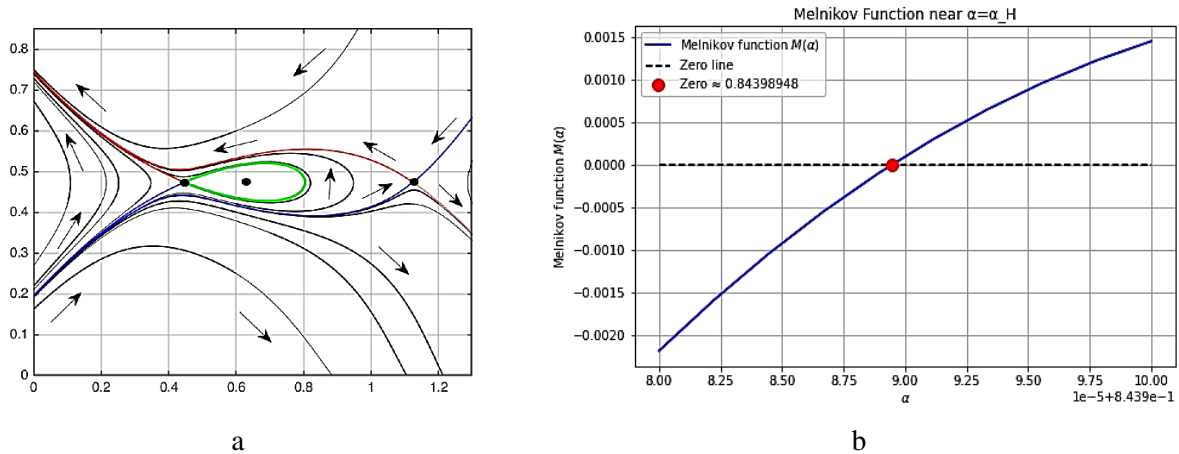

a                                b

**Figure 13**: For the parameters $\gamma_1 = 0.92$, $\gamma_2 = 8.33$, $\mu = 0.8$, $\beta = 0.4$ we show (a) the emergence of Homoclinic bifurcation for $\alpha_H \approx 0.844$. (b) by assessing the Melnikov integral from a range of various $\alpha$'s as the model's bifurcation parameter, we show where it crosses zero.

As we laid out earlier, changes in parameter values can force players into a regime of multiple equilibria, leading to severe consequences that would signal their collapse. Therefore, in combat scenarios, it is vital for players to avoid this situation. For example, implementing targeted adjustments to the collateral damage function can help obviate the need to consider costly contingencies. A key question arises: is it possible to eliminate multistability from the system?

As demonstrated in Figure 3, a decrease in the attrition rate $\gamma_1$ can both induce and remove multistability. We now aim to analyze this issue through the lens of a codimension-2 bifurcation framework, focusing on how changes in the mathematical structure of $\theta$ influence system dynamics, in tandem with a changing $\gamma_1$, can result in the elimination of the Hopf bifurcation zone. Simply put, since $\theta(C)$ is dealing with the local population[56], a possible change would turn three equilibria into one, thus adding to the complexity of the system.

The key to understanding the importance of the change of $(\gamma_1, \eta)$ is that some constraints, like regional and international pressures, may change $\gamma_1$ and $\eta$ and eliminate the Hopf bifurcation zone. The new technological instruments have heralded a new era of tensions between various players. Given the fast rate of dissemination of information[57], the effect of $\eta$ in the models and its complexity should be taken seriously. More importantly, however, it may bring the system's equilibria closer to each other, thus giving rise to the occurrence of heteroclinic bifurcation in tandem with varying $\gamma_1$.

---

[56] Generally speaking, it can represent various factors strengthening insurgents.

[57] Rudner, Martin. ""Electronic Jihad": The Internet as al-Qaeda's catalyst for global terror." In *Violent Extremism Online*, pp. 8-24. Routledge, 2016.; Parvez, Saimum. "How Do Terrorists Use the Internet for Recruitment?." In Parvez, Saimum, and Mohammad Sajjadur Rahman, eds. *The Politics of Terrorism and Counterterrorism in Bangladesh*. Routledge, 2022.; See Weissmann, Mikael. "Urban warfare: challenges of military operations on tomorrow's battlefield in Weissmann, Mikael, and Niklas Nilsson (Eds) . *Advanced Land Warfare: Tactics and Operations*. Oxford University Press, 2023, pp 146-147

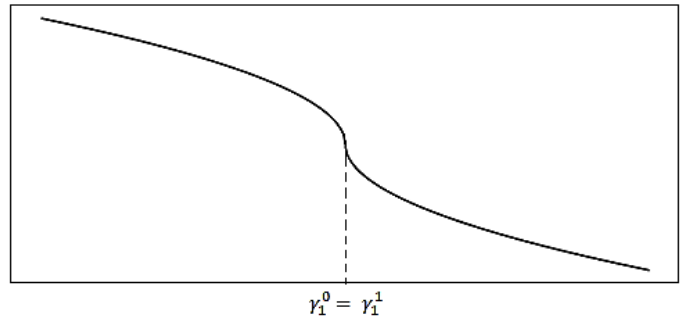

$\gamma_1^0 = \gamma_1^1$

**Figure 14** : the occurrence of the Cusp Bifurcation as a result of change in $\gamma_1$ and $\eta$ ,as a parameter of $\theta(C)$. The difference between this figure and figure 4 is that a two dimensional change has had the red area eliminated

## Extension

A retrospective analysis of the Lanchester model of warfare alludes to an important question arising from real-world applications. How adding more players[58] shapes the conflict and impacts the dynamics of our conflict, which underwent both local and global bifurcations. Generally, the idea of alliance and rivalry within groups depends on many conditions, which reflect demographic, religious, and regional issues. In the case of Afghanistan, after the withdrawal of soviet forces, we saw a chain of different alliances and rivalries among the nation's ethnic groups, from a united minority in 1992 to the formation of an alliance against the Taliban in 1996[59], shows the dynamics of friendship and animosity in a fragile state. The same argument applies to Syria. Before sketching a possible extension of our model, it is important to address how the concept of collateral damage is distributed among different players, namely $I = (I_1, I_2, ... I_n)$ and how G might respond to the likely turf or alliances within I.

In fact, damages and civilian casualties, given the problem of attribution, can have dual effects on insurgents. While insurgents and some aligned factions may use strategies to maximize civilian harm by using civilian structures, some factions or defectors may apportion blame on insurgents and their allies for the harm inflicted on the targeted area. As we mentioned earlier, one of the most important aspects of the Lanchester model is to pay extra attention to the dynamics of population attitude and its effect on the battle between $G$ and $I$ .Our sophisticated collateral damage will pave the way for possible deviations from the conventional wisdom on the relationship between un-aimed attacks and the population's role in the conflict. In fact, our formalism for the suggested damage function will give credence to this deviation; by introducing $\gamma_2 I (1 - I)$ , We narrowed down our focus to the inevitability of damage in high-intensity conflicts like urban warfare and the likelihood of targeting a mixture of insurgents and population in a specific area. This mixture adds another complexity to the conflict and may have multiple scenarios with far-reaching impact on both players. So far, we asserted that a lack of intelligence can turn harmed civilians into insurgents seeking vengeance. Now, we put forward another possibility that lays the groundwork for causing drawbacks for insurgents on the horizon of conflict. The term" double-edged sword" attacks, which provided an exegesis for the pros and cons of insurgency for the government, can be applied to insurgents who may be stabbed in the back by a percentage of infuriated civilians. A retrospective analysis of insurgency illustrates that the problem of attribution in battles happening in the presence of civilian

[58] Or variables in general
[59] Christia, Fotini. *Alliance formation in civil wars*. Cambridge University Press, 2012., pp 85-89

populations is much more complicated than we might think.[60]. Moreover, an increase in insurgent attacks and retaliation from the side of the government can embitter the relationship between insurgents and some civilians, thus adding to the latter's anger[61]. For instance, according to a statistical analysis, there is a difference between recipients and observers of attacks regarding the attribution of blame. [62]. In some cases, the civilian uprising against rebels and insurgents plays a vital role in governments' counterinsurgency strategies. [63] Meanwhile, the comprehensive category of curves will get into the act to appraise how different rival insurgent groups benefit from or are harmed by collateral damage structures.[64] Besides the presence of local communities blaming the insurgents for their policies and actions , the complexity of battlegrounds in the 21st century, like a vast population, will make the emergence of local communities standing up to insurgents [65] more likely. Therefore, having a proper understanding of the dynamics of angry populations and the strategies both governments and insurgents adopt to maximize their goals is a sine qua non[66]

In the case of a multi-agent insurgent militia, higher-dimensional invariant objects may be observed. For example suppose $I = (I_1, I_2)$. If based on the setting of the problem, the fixed points stay on $G$ −axis, then in the three dimensional $(G, I_1, I_2)$ − phase space, a three-dimensional oval shape similar to the rotation of Figure 12 around the $G$ −axis may occur. Then, the formalism of the twist map theorem[67] will be required to describe the whole dynamics. See Figure 15 for a 3-d graphical model schematically. However, the exact details require a comprehensive study , which is beyond the scope of this paper. Another important issue in higher-dimensional analysis is the complexity of the control problem. Suppose we want to control the state from a present point at a given level set of the invariant surface $\partial\Omega$ to an admissible targeting point at another level set of the invariant surface $\partial\Omega$. A control orbit confined to $\partial\Omega$, is depicted in Figure 16. This problem will recall the control set techniques in the dynamics of control[68]. Control sets are disjoint, approximately completely controllable domains that can be detected using compact invariant objects of various bifurcations, such as Hopf and homoclinic bifurcations.

[60] Condra, Luke N., and Jacob N. Shapiro. "Who takes the blame? The strategic effects of collateral damage." *American journal of political science* 56, no. 1 (2012): 167-187.

[61] Lyall, Jason. "Does indiscriminate violence incite insurgent attacks? Evidence from Chechnya." *Journal of Conflict Resolution* 53, no. 3 (2009): 331-362. PP 337

[62] Pechenkina, Anna O., Andrew W. Bausch, and Kiron K. Skinner. "How do civilians attribute blame for state indiscriminate violence?." *Journal of Peace Research* 56, no. 4 (2019): 545-558. Generally speaking, mass killing and other strategies followed by insurgents like using people has human shields can decouple population.

[63] For example, the Iraqi tribes who took up arms to counter jihadists helped American forces. See Henriksen, Thomas H. *America's wars: interventions, regime change, and insurgencies after the Cold War*. Cambridge University Press, 2022. P 10; In some battles, we can find the trace of paramilitary groups supporting governments See Arsene Kabore and Sam Mednick, Burkina Faso rights group alleges 28 dead in ethnic killings, Associated Press, Jan 3 , 2023, https://apnews.com/article/politics-burkina-faso-crime-military-and-defense-violence-02aa75e6bcb471115186210f8d1217d9

[64] Walter, Barbara F. "The extremist's advantage in civil wars." *International security* 42, no. 2 (2017): 7-39., https://doi.org/10.1162/ISEC_a_00292

**[65]** Or governments

[66] In this model, we do not aim to discuss 3 players but our goal is to mention the collaboration between harmed population and government's support. For 3D Lanchester models see Kress, Moshe, Jonathan P. Caulkins, Gustav Feichtinger, Dieter Grass, and Andrea Seidl. "Lanchester model for three-way combat." *European Journal of Operational Research* 264, no. 1 (2018): 46-54., https://doi.org/10.1016/j.ejor.2017.07.026

[67] Guckenheimer, John, and Philip Holmes. *Nonlinear oscillations, dynamical systems, and bifurcations of vector fields,* Springer, 2013. P 219.

[68] Colonius, Fritz, Kliemann, Wolfgang, *Dynamics of control*, Birkhauser, 2000.

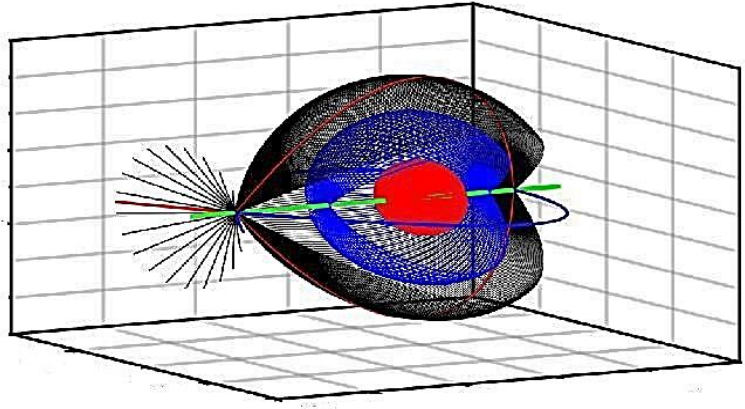

**Figure 15:** A 3-d graphical model for $(G, I_1, I_2)$ −dynamics, schematically. Compare with Figure6

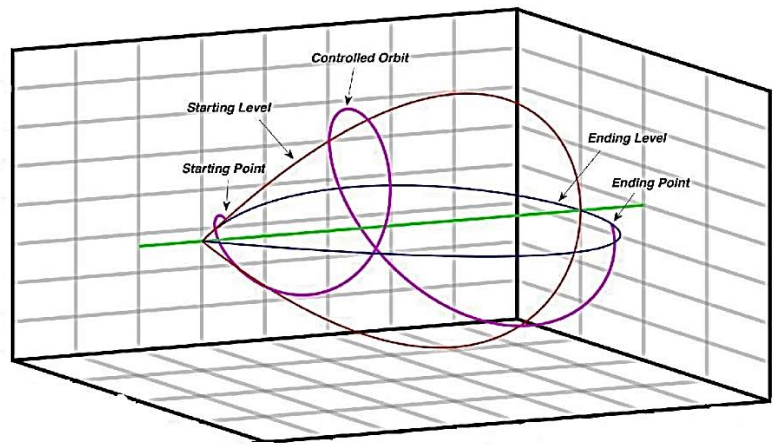

*Figure 16: A 3-d graphical scheme of a twisted controlled orbit in $(G, I_1, I_2)$ −dynamics, schematically*

Meanwhile, in a strategic interaction, the training of a portion of harmed civilians should be optimized in order to minimize pressure on G. Furthermore, it is worth noting to discuss the mathematical characteristics of the infuriated population curve in order to figure out the degree to which these individuals are supporting the government's agenda. It goes without saying that the sense of anger, despair, and radicalization is not unlimited, and there is a nuanced boundary that signals changes in opinions regarding governments' actions. For instance, tough measures and indiscriminate violence can cause the shrinkage of this support for the government's action. Therefore, the degree of brutality has a mixed impact on the fissure within the targeted area; notwithstanding its benefits for the government that may separate insurgents from a portion of the population, too much violence could diminish the alternative support, thus tarnishing the horizon for ending the conflict.

## Discussion

The main goal of this analysis was to show the sophisticated versions of collateral damage functions and their repercussions. Facing various conflicts with unique characteristics will necessitate us to shed special light on the typology of collateral damage functions that drive the conflict and give rise to the occurrence of local and global bifurcations in our system. We showed two main types of collateral damage functions. The first category explained not necessarily convex functions that featured being saturated after a specific threshold; in this form, the curve may be less sensitive to minimal damage. Meanwhile, after passing the threshold, the level of response to damage will remain relatively constant. The second category mentioned the possibility of having non-monotonic curves. Excess violence may cause the affected population to turn away from insurgents, thus slowing down their recruitment or logistics.[69] In our work, we made a notable qualification of the damage function and assessed the possibility of a non-monotonic relationship between I and C according to which, ranging from zero to 0.5, increase in the proportion of insurgents will increase the damage; the key to understand this reality stemmed largely from the fact that $I = 0$ implies no non-ideological drive for insurgency in this battle. In the meantime, we showed the effect of effective targeting of insurgents diffused in a specific area can be higher than the rate of attrition, which is understandable in complex battlegrounds.

The second part of our analysis involved conducting a bifurcation analysis of the dynamical system to uncover its underlying complexities. While the notion of limit cycle and periodic behavior in the $(G, I)$ conflict was widespread and well-known, the ramifications of a collapse will open a new door to the insurgency scholarship. The existence of multistability, which stems from the categorical analysis of collateral damage functions, forces players to play a dangerous game of periodic tradeoffs. Simply arguing, a change in the amplitude of the cycle due to a change in combat parameters, which is widespread in conflicts, can terminate the cycle and bring an end to the conflict. A retrospective analysis of wars and insurgencies from Vietnam to Afghanistan and Syria places a special premium on the importance of global bifurcations in the Lanchester model of warfare. More importantly, acquiring an analytic view of homoclinic bifurcations[70] will enable us to clearly illustrate the existence of homoclinic orbits by using the Melnikov function. The Hamiltonian version of our model, despite its idiosyncrasies, resembles the behavior of the original model in terms of equilibrium analysis and helps us study the system in the presence of a perturbation, which was construed as the effect of lacking intelligence[71]. In the tradeoff between G and I. Then, by pointing to the Cusp bifurcation, we paved the way for codimension two assessment of our model. Finally, by bringing up the possible extension of our model, we added a new layer of applicability and complexity to the Lanchester model of warfare that points to how different factions perceive collateral damages and attribute blame.

---

[69] It is worthwhile to note that we exclude regional or international barriers in counterinsurgency, which may overturn the decline in $\theta$ to some extent.

[70] Keep in mind that heteroclinic bifurcation, provided that the three equilibria are close to each other, can add more complexity to our model because it would explain the circumstances under which both players, $G$ and $I$ are on the verge of collapse, and a minuscule variation in initial conditions or parameters can have a huge and probably irreversible impact on them.

[71] Or $1 - \mu$

Due to the strategic nature of this interaction, one can raise the question of the control analysis of this model as a prelude to the game-theoretic version of the sophisticated Lanchester model. The occurrence of a Homoclinic bifurcation and its draconian consequences for players will compel players or external actors to control the dynamics of the conflict. A control-set version of our model will set the stage to study controllability in near-collapse areas, which is vital for players in counterinsurgencies. By focusing on the parameters of the collateral damage function, analysts and experts can conduct a data-driven analysis of curves to comprehensively categorize them based on grievance, sophisticated battle environment, and technological factors. More interestingly, in the post-collision circumstances, players can deviate from rationality and take more risks, which makes the connection between our model and relaxed rationality models, including bounded rationality and prospect theory near global bifurcations.